\documentclass{llncs}

\usepackage[caption=false,font=footnotesize]{subfig}
\usepackage{tabularx}
\usepackage{paralist}
\usepackage{balance}
\usepackage{todonotes} \presetkeys{todonotes}{inline}{}
\usepackage[subpreambles]{standalone}

\usepackage[T1]{fontenc}
\usepackage[utf8]{inputenc}
\usepackage{booktabs}
\usepackage{xspace}
\usepackage{graphicx}
\usepackage{paralist}
\definecolor{cite}{rgb}{0,0,0}
\definecolor{link}{rgb}{0,0,0}
\definecolor{url}{rgb}{0,0,0}
\usepackage{hyperref}
\hypersetup{
  colorlinks,
  citecolor=cite,
  linkcolor=link,
  urlcolor=url}
\usepackage{amsmath}

\newcommand{\eg}{e.g.\@\xspace}
\newcommand{\ie}{i.e.\@\xspace}

\usepackage{tikz}
\usepackage{algorithm,algorithmic}

\newcommand{\TopicFocus}{C-F\xspace}
\newcommand{\TempFocus}{T-F\xspace}
\newcommand{\TopicTempFocus}{CT-F\xspace}

\newcommand{\CS}{Collection Specification\xspace}
\newcommand{\events}{28\xspace}

\newcommand{\timedecay}{\ensuremath{\gamma}}

\usepackage{amsthm}
\theoremstyle{definition}
\newtheorem{mydef}{Definition}

\tikzset{font=\footnotesize}

\begin{document}

 \title{Extracting Event-Centric Document Collections from 
 Large-Scale Web Archives}

\author{Gerhard Gossen \and Elena Demidova \and Thomas Risse}
\institute{L3S Research Center, Leibniz Universität Hannover\\
\email{\{gossen, demidova, risse\}@l3s.de}}

\maketitle

\begin{abstract}
Web archives are typically very broad in scope and extremely large in scale.
This makes data analysis appear daunting, especially for non-computer scientists.
These collections constitute an increasingly important source for researchers in the social sciences, the historical sciences and journalists interested in studying past events.
However, there are currently no access methods that help users to efficiently access information, in particular about specific events, beyond the retrieval of individual disconnected documents.
Therefore we propose a novel method to extract event-centric document collections from large scale Web archives.
This method relies on a specialized focused extraction algorithm. Our experiments on the German Web archive (covering a time period of 19 years) demonstrate that our method enables the extraction of event-centric collections for different event types.
\end{abstract}

\setdefaultleftmargin{1em}{2em}{1.87em}{1.7em}{1em}{1em}
\setdefaultitem{$\circ$}{\normalfont\bfseries \textendash}{\textasteriskcentered}{\textperiodcentered}

\section{Introduction}
\label{sec:introduction}

Web archives created by the Internet Archive\footnote{https://archive.org} (IA), national libraries and other archiving services contain large amounts of information collected for a time period of over twenty years~\cite{Costa2016}.
These archives constitute a valuable source for research in many disciplines, including the digital humanities, the historical sciences and journalism by offering a unique possibility to look into past events and their representation on the Web.
They can enable a better understanding
of past events and offer a lot of novel research directions for these disciplines.

Most Web archive services aim to capture the entire Web (IA) or national top-level domains (national libraries) and are therefore very broad in their scope.
Consequently they are also very diverse regarding the topics they contain and the time intervals they cover.
Due to the large size and the broad scope it is difficult for interested researchers to locate relevant information in the archives as search facilities are very limited compared to the live Web.

In previous work~\cite{icrawlRequirements,gossen2016websci} we have argued that these users
are typically interested in studying smaller and more focused event-centric collections of documents contained in a Web archive.
Such collections can reflect specific events such as elections, sports tournaments or natural disasters, for example the Fukushima nuclear disaster in 2011, the German federal election in 2009 or the FIFA World Cup 2006, especially in regard to their media coverage and public reactions.

Archive services such as Archive-IT\footnote{\url{https://archive-it.org/}} collect documents around specific events.
These \emph{special collections} are however defined and crawled on an individual basis,
such that users are restricted to the collections that exist and their selected scope.
Other existing access methods to temporal Web collections do not support creating ad-hoc collections,
often forcing users to create their own corpora manually.
Currently, access to large-scale Web archives is limited to browsing of individual Web pages through browser-based tools such as the Wayback machine\footnote{\url{http://netpreserve.org/openwayback}}, 
or initial support for  keyword-based access\footnote{\url{https://blog.archive.org/2016/10/24/beta-wayback-machine-now-with-site-search/}}.
However, these access methods are not sufficient for several reasons.
First, the Wayback machine requires the user to already know the URL of the document.
Second, full-text indexing of large-scale archived collections incurs high processing and storage costs.
Third, such indexes only allow retrieval of individual disconnected documents.
Instead, automatic methods are needed that can extract collections of
documents related to a particular event of user interest. These
collections need to preserve the original link structure to achieve a high
degree of authenticity and enable the application of analytical methods
on the relevant parts of the Web archive~\cite{gossen2016websci}.

In this paper, we present a starting point for tackling the novel problem of extracting topically and
temporally coherent, interlinked event-centric document collections from
large-scale and broad scope Web archives.
The key contributions of this paper are: 
\begin{inparaenum}[(1)]
\item a definition of a \emph{\CS} that describes the temporal and topical scope of the collection to be extracted
and gives the user intuitive but powerful options to control the data collection process; and
\item a \emph{focused crawling-based extraction method} for Web archives to
 create event-centric collections without requiring any full-text indexes.
We evaluate our approach in a local environment using file system crawling.
However, our approach can easily be used across Web archives using existing access methods.
We make our source code and evaluation data available to encourage further research\footnote{\url{https://github.com/gerhardgossen/archive-recrawling}}.
\end{inparaenum}

\section{Related Work}
\label{sec:related-work}

Our method is related to crawling methods for creating Web
Archives 
(e.g. \cite{jiang_focus_2013,PereiraMCM14}),
as well as to methods
for temporal information retrieval \cite{Costa:2014:LTR:2600428.2609619}.

The collection of Web documents from the live Web
for retrieval and archiving purposes is usually performed using Web crawlers.
Crawling methods that aim to create broad scope collections for
search and archiving purposes 
intend to capture as much of the Web as possible.
An example of a web-scale archiving crawler currently used by
the Internet Archive is Heritrix \cite{mohr04heritrix}.
In contrast, \emph{focused crawling} \cite{chakrabarti99:focus} aims to only
collect pages that are related to a specific topic.
Focused crawlers \cite{Aggarwal:2001:ICW:371920.371955,Pant2005}
learn a model of the topic and follow links only if they are expected to match that topic, \eg based on the page containing the link.
This follows the obervation that relevant documents will preferentially 
link to other relevant documents (``topical locality'' \cite{Aggarwal:2001:ICW:371920.371955}).
Extensions of this model use ontologies to incorporate semantic knowledge
into the matching
process \cite{Ehrig2003,Dong13}, `tunnel' between disjoint page
clusters \cite{bergmark_focused_2002,jialun_qin_building_2004} or learn navigation
structures necessary to find relevant pages~\cite{Diligenti2000,jiang_focus_2013}.
In time-aware focused crawling \cite{PereiraMCM14} the
document or event time is used as the primary focusing criterion.
In event-based crawling~\cite{Farag2017} events are described using an event model that incorporates event location and date.
Here Web page relevance is computed as a weighted average of content, location and date similarity.
As location extraction increases the overall complexity of the process,
we focus on the content and time-based features. Freshness as a specific aspect of
temporal relevance has been addressed in the context
of joint crawling of the Web and Social media sites~\cite{gossen2015icrawl}
where URLs present in Social media posts are used as entry points to recently published content on the Web.
In summary, most existing approaches to focused Web crawling consider
the topical and temporal relevance in isolation and do not address the
problem of jointly finding temporally and topically relevant content.
Furthermore, whereas existing approaches operate on the live Web, we are the
first to apply focused crawling techniques to existing Web archives.

The notion of temporal relevance has
also been explored in the area of temporal information retrieval.
Existing ranking methods have been extended to rank documents based on their creation time
\cite{Costa:2014:LTR:2600428.2609619} or to diversify search
results over relevant time periods
\cite{berberich2013temporal}.
Contemporary search engines also rank documents based on their freshness (estimated based on their crawling history)~\cite{Dong:2010:TRR:1718487.1718490}.
Similarly, time information has been combined with the hypertext link
graph to detect the most relevant documents for a given
query \cite{Nguyen:2015:TRW:2766462.2767832}.
These approaches depend on full-text or graph indexes and therefore have
a high up-front computational and index storage cost.
Moreover, these approaches only allow retrieval of individual disconnected
documents and do not preserve the link structure.
In contrast, our method allows on demand extraction of interlinked
event-centric collections without requiring any additional indexes on the
archive.

\begin{table}[t]\small
  \caption{Examples of temporal event characteristics.}
  \label{tab:event-characteristics}
  \centering
  \begin{tabular}{lllll}
    \toprule
    Event              &Type& Duration \ & Lead time & Cool-down time  \\
    \midrule
    Olympic games     & recurring & 2 weeks   & weeks   & days \\
    Federal election   & recurring & 1 day     & months  & weeks \\
    Fukushima accident \  & non-recurring \  & 1 week    & ---     & months \\
    Snowden leaks      & non-recurring & 1 day     & ---     & years \\
    \bottomrule
  \end{tabular}
\end{table}

\section{Event-Centric Collections}
\label{sec:problem}

Events are typically characterized through a certain date or a
time interval such as the date of an accident or the duration of a
tournament.
Here the event time interval is clearly defined.
Nevertheless, event-related documents also appear outside of this time interval.
For planned and in particular regularly recurring events such as sports
competitions or elections, relevant documents often appear in advance of
the actual begin of the event during the event \textit{lead time},
and are still published after the event completion during the
\textit{cool-down time}.
For unexpected non-recurring events such as natural disasters,
event-related documents are published from the start of the
event onward, \ie there is no lead time and the relevant
documents appear during the cool-down time of the event.
The duration of the lead time as well as the duration of the cool-down time
depend on the specific event (see Table~\ref{tab:event-characteristics}).

Given an event of user interest and a large-scale broad-scope Web archive, our goal is to
generate an interlinked collection of documents relevant to this event.
The scope of the target collection is
defined in the \textbf{\CS}:

\begin{mydef}[\CS]
\label{def:crawl-spec}
The \CS defines the topical and the temporal scope of an event-centric
collection using:
\begin{compactitem}
\item Topical Scope:
\begin{compactitem}
  \item one or more topical reference documents (e.g. from the Web);
  \item zero or more representative \emph{keywords}.
\end{compactitem}
\item Temporal Scope:
\begin{compactitem}
    \item time span of the event (including the start and end dates) $T_e=[t_e^s , t_e^e]$;
    \item time duration of the lead time ($T_l$) and the cool-down
    time ($T_r$).
\end{compactitem}
\end{compactitem}

\noindent The \CS may be extended to include additional scopes, for example domain black and white lists as used by existing crawlers.
\end{mydef}

Given the \CS, our goal is to create a collection containing the Web documents temporally and topically relevant to  this specification.
In the following we
propose
a focused extraction method that prioritizes URLs during
the crawling process according to the \CS and generates interlinked event-centric
collections.

\section{Event-centric Collection Extraction}
\label{sec:archive-re-crawling}

Our goal is to efficiently extract an event-centric interlinked collection of a manageable size from a large scale Web archive.
A na\"ive approach is to iterate through all documents
and check their relevance with respect to the \CS using
an automatic method. However, this is computationally expensive and does not scale 
to Web archives spanning tens or hundreds of terabytes.
While a full-text index could reduce the iteration cost, 
it requires high up-front computational and index storage resources
and extensive post-filtering of the many
near-identical document versions contained in the Web archive~\cite{jackson2016}.
Furthermore, such an index can only be used to retrieve individual documents,
where we want to extract interlinked document collections.

We propose an alternative approach that uses the hypertext characteristics of the
archived documents by adapting focused Web crawling.
A Web crawler collects documents by recursively following the links from a Web document to other documents, 
starting from an initial set of \emph{seed URLs}.
A focused Web crawler improves the relevance of the
resulting collection by following only links to the documents predicted to be
relevant.
We therefore extend the \CS to include the seed URLs required for the crawling process:
\begin{mydef}[Crawl-based \CS]
  A Crawl-based \CS contains a \CS (Definition~\ref{def:crawl-spec}) and a non-empty set of URLs,
  which are contained in the archive and refer to relevant documents.
\end{mydef}

The Crawl-based \CS is created by the user.
Semi-auto\-matic approaches 
include the use of Web search
engines to select seed URLs~\cite{gossen2015wizard}.

\newcommand{\from}{\leftarrow}
\renewcommand{\algorithmicrequire}{\textbf{Input:}}
\renewcommand{\algorithmicensure}{\textbf{Output:}}
\begin{algorithm}[bt]
  \caption{Event-centric Collection Extraction}
  \label{alg:recrawling}
  \begin{algorithmic}
    \REQUIRE \CS $CS$, $targetSize$
    \ENSURE Document collection $c$, excluded URLs $missing$
    \STATE $q \from $ priorityQueue(seedUrls($CS$));  $c \from \{\}$;  $missing \from \{\}$
    \WHILE{\NOT isEmpty($q$) \AND $|c| < targetSize$}
      \STATE $url  \from pop(q)$
      \STATE $v \from resolveSnapshots(url, CS)$ \COMMENT{Find all snapshots of $url$ in $c$}
      \IF{$v = \emptyset$}
         \STATE $missing \from missing \cup \{url\}$
      \ELSE
         \STATE $v_i \from selectSnapshot(CS, v)$
         \STATE $c \from c \cup \{v_i\}$
         \STATE $out \from extractOutlinks(v_i) - seenUrls$ \COMMENT{$seenUrls = c \cup missing$}
         \STATE $insert(q, out, relevance(v_i)$ \COMMENT{Insert outlinks into queue according to relevance}
      \ENDIF
    \ENDWHILE
  \end{algorithmic}
\end{algorithm}

We adapt the focused crawling algorithm as shown in 
Algorithm~\ref{alg:recrawling}
by including steps to resolve snapshots and select the best among them.
URLs extracted from collected documents are prioritized in the crawler queue during the focused
crawl using the relevance function 
defined in Section \ref{sec:scoring}.

\section{Relevance Estimation}
\label{sec:scoring}

We need  to prioritize the URLs during the focused crawl
to effectively extract event-centric collections
based on a \emph{relevance function}.
We use a linear combination of the temporal and topical
relevance (TTR) to estimate the relevance
of a Web document $d$ with respect to the \CS $CS$:
\begin{equation}
  \label{eq:TTR}
  \begin{split}
  \text{TTR}(d, CS) =
  &\ \alpha\times\text{TopicR} (d, CS )
  + (1-\alpha)\times\text{TempR}(d, CS),
    \end{split}
\end{equation}
\noindent where $TempR$ and $TopicR$ are the temporal and topical relevance of $d$ to $CS$,
and $\alpha \in [0,1]$ is the parameter to trade off between the
topical and temporal relevance.
$\alpha = 1$ results in a standard topically focused crawler, whereas values closer to $0$ increase the weight of the temporal dimension.
In our setting we consider $TempR$ and $TopicR$ to be equally important,
therefore we use $\alpha = 0.5$, but we will in future work investigate the influence of this parameter in detail.

\subsection{Temporal Relevance}
\label{sec:temporal_relevance}

As described in Section~\ref{sec:problem}, event-related documents are
published not only during the event time interval, but also before and after.
Consequently, we need to estimate the relevance of a document based on the
\CS and a time point associated with the Web document
(e.g. the creation, last modification or capture date).
We define this \emph{Temporal Relevance Function} as follows:
\begin{mydef}[Temporal Relevance Function]
\label{def:temporal_relevance_function}
Given a time point $t_d$
associated with the Web document $d$
and the event time interval $T_e = [t_e^s, t_e^e]$,
the function $f(t_d, t_e) \to [0,1]$ is a
temporal relevance function iff (a) $f(t_d, t_e) = 1 \Rightarrow t_d \in t_e$ and
(b) $f$ is monotonically non-decreasing
in $(-\infty, t_e^s)$ and monotonically non-increasing in $(t_e^e,+\infty)$.
\end{mydef}

We assume that in general the relevance of documents decreases rapidly as the
distance to the event increases and therefore define a temporal relevance
function based on the exponential decay
function (similar to \cite{Kanhabua:2011:CTR:2009916.2010147}):
\begin{equation}
  \label{eq:TempR-single-link}
  \text{TempR} (t_d, t_e) =
    \begin{cases}
     1 & \text{if $t_e^s \le t_d \le t_e^e$,} \\
     e^{-\Delta t/\timedecay_l} & \text{if $t_d < t_e^s$,}\\
     e^{-\Delta t/\timedecay_r} & \text{if $t_d > t_e^e$,}\\
    \end{cases}
\end{equation}

\noindent where $ \Delta t $ is the time difference between the document time
point $t_d$
and the nearest end of the reference time interval $T_e$, and
$\timedecay_l$ and $\timedecay_r$ are \emph{time decay factors}.
The time decay factors determine how fast the value of this function
decreases by giving the $\Delta t$ at which the relevance has dropped to $0.5$.
We use the expected duration of the lead and the cool-down time as the
time decay factors $\timedecay_l$ and $\timedecay_r$.
For events with no lead time (\eg accidents) we set $\timedecay_l=0$.

The document time point can be estimated using the date discussed in the document.
This would give the most accurate relevance value, especially for documents that describe the event after some time has passed (\eg at the one year anniversary),
but is computationally expensive and highly heuristic.
Therefore we extract the document publication time, which is often explicitly contained in the document metadata or content.
If no publication time is available, we use the crawl time as a fallback.

\subsection{Topical Relevance Estimation}
\label{sec:focussing-relevance}

The topical relevance of Web documents with respect to the \CS is
estimated by computing the similarity of
the textual content of Web documents to the topical
scope of the \CS (similar to \cite{pant2004crawling}).

The topical scope is specified primarily through a set
of \emph{reference documents} that describe the event (\eg
as Wikipe\-dia pages or newspaper articles).
When these documents have an ambiguous topic or the scope
should be narrowed down further, keywords can be provided to clarify the topical intent.
Together this allows an intuitive yet powerful topical specification.
We represent the topical scope as a term vector, called the \textit{reference vector}, to enable automatic relevance estimation with respect to the topic.
To construct the reference vector we tokenize and stem the text of the reference documents
and remove stop words using the language-specific analyzers of
Apache Lucene\footnote{\url{http://lucene.apache.org/core/}}.
As previous work has shown bigrams to be effective for crawl
focusing~\cite{Laranjeira2014},
we use term unigrams and bigrams.
Each term is weighted using its frequency (TF) and its inverse document frequency (IDF).
IDF scores are based on the frequencies of the last 25 years of the Google Books
NGram datasets\footnote{Code available at:
\url{https://github.com/gerhardgossen/dictionary-creator/}}.

The weights of terms explicitly given as \CS keywords
are boosted.
This helps to shift the reference vector towards the expected interpretation.
To perform boosting, we check the overlap of each term with the
user-defined keywords, as terms (in the case of bigrams) can contain multiple
tokens.
Based on whether there is a full or partial overlap,
we assign a \emph{term weight} $tw_t$ to the term $t$ in the document vector.
In our evaluation, we experimentally set the values
for full, partial and no overlap 
to $2$, $1.5$ and $1$, respectively.

Finally, the topical relevance of a document is the cosine similarity between the reference vector and a document vector computed using the same method.

\section{Web Archive and Platform}
\label{sec:dataset}

Our Web Archive contains all Web
pages from the \texttt{.de} top-level domain as captured by the Internet Archive until 2013.
In this paper we only consider HTML documents with a HTTP status code of 200.
This archive has a size of about 30 TB and contains 4.05 billion captures of 1 billion URLs,
covering a time period from December 1994 to September 2013.

We manually defined \events events to be extracted from the Web archive, focusing on events that are
likely to be represented in the archive:
The selected events fall within the time period of the archive and have a strong connection
to Germany, either because the event happened in Germany or was in the focus
of public attention.
We balanced singular events like the Fukushima nuclear accident
and recurring events like federal elections.
To create the \CS for each event we selected one or more pages from the German Wikipedia that provide the
topical scope of the event.
We also defined a start and end date, as well as an estimate for the duration of the
event lead and cool-down time.
The outgoing links of the Wikipedia pages were extracted and used as seed URLs. 

All experiments were conducted on a Hadoop cluster.
This cluster has 25 worker and 2 master nodes with in total 296 CPU cores.
The worker nodes provide in total 1.37 TB of RAM and 1 PB of hard disk capacity.
All data is stored in the standard ARC/WARC formats 
and available to all worker nodes.

\subsection{Crawler Implementation}
\label{sec:implementation}

As mentioned in Section \ref{sec:archive-re-crawling}, the architecture
of the archive crawler can be simpler than that of a standard Web crawler
because it can access the data of the Web archive locally.
As our data is stored as WARC files in a Hadoop filesystem,
we implemented the crawler as a multi-thread process running on Hadoop YARN.

WARC files are unordered collections of documents,
therefore a lookup table is necessary
to find the location of the document snapshots for a given URL.
By using Apache HBase for this table we can look up
URLs in 1-5 milliseconds. While typically CDX files are used as a
lookup method for WARC files, our preliminary experiments
showed that this method is considerably faster.

The crawler queue is stored in a file-based queue based on the Mercator
architecture \cite{heydon1999mercator}, which offers prioritisation of
URLs and is fast enough for our purposes.
Each retrieved document is analysed according to the relevance function
described in Section~\ref{sec:scoring}.
The URLs of all outgoing links of that document are inserted into the crawler queue according
to the calculated relevance score.

As the Web archive covers a long time period, many documents have been
crawled multiple times. To choose among the available versions,
we observe that later versions typically have the same content
but may have changes in \eg navigation menus
and thus do not represent the document in its original form.
Therefore we use the following heuristic:
If multiple versions are available that were crawled during
the event timespan, we pick the earliest.
Otherwise, we use the version that was crawled closest to the event timespan.
Future work will investigate further methods to select the most relevant version(s).

\section{Evaluation}
\label{sec:evaluation}

The goal of the evaluation is to assess the precision of the proposed
extraction method in light of
different event types and to better understand the influence of this
method on the quality of the resulting event-centric collections.
We compare our combined relevance function with two baselines
that use state-of-the-art relevance functions, each taking only one relevance
dimension into account, topical (\TopicFocus, cf.~\cite{pant2004crawling}) or temporal (\TempFocus, cf.~\cite{PereiraMCM14}).
We also use an unfocused crawl that does not use any relevance estimates as an additional baseline.

\begin{figure}[t]
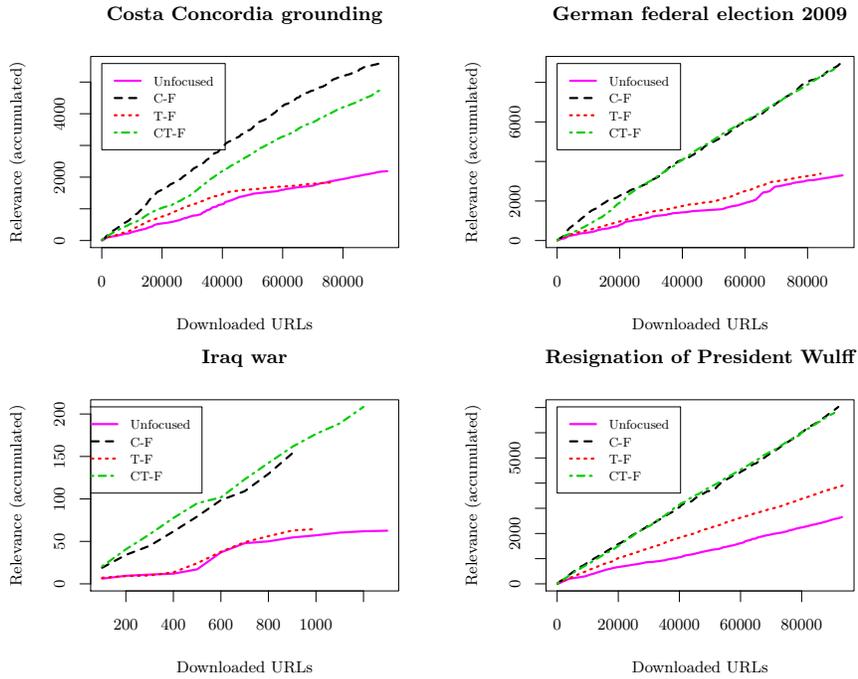

  \vspace{-\baselineskip}
  \centering
  \scalebox{.68}{\input{figures/costa}}
  \scalebox{.68}{\input{figures/election2009}}

  \vspace{-2em}
  \scalebox{.68}{
\begin{tikzpicture}[x=1pt,y=1pt]
\definecolor{fillColor}{RGB}{255,255,255}
\path[use as bounding box,fill=fillColor,fill opacity=0.00] (0,0) rectangle (245.72,216.81);
\begin{scope}
\path[clip] ( 49.20, 61.20) rectangle (220.52,167.61);
\definecolor{drawColor}{RGB}{255,0,255}

\path[draw=drawColor,line width= 1.2pt,line join=round,line cap=round] ( 55.55, 68.03) --
	( 68.76, 69.62) --
	( 81.98, 70.29) --
	( 95.20, 70.82) --
	(108.42, 73.21) --
	(121.64, 82.67) --
	(134.86, 87.75) --
	(148.08, 88.80) --
	(161.30, 90.95) --
	(174.52, 92.16) --
	(187.73, 93.67) --
	(200.95, 94.41) --
	(214.17, 94.73);
\end{scope}
\begin{scope}
\path[clip] (  0.00,  0.00) rectangle (245.72,216.81);
\definecolor{drawColor}{RGB}{0,0,0}

\path[draw=drawColor,line width= 0.4pt,line join=round,line cap=round] ( 68.76, 61.20) -- (200.95, 61.20);

\path[draw=drawColor,line width= 0.4pt,line join=round,line cap=round] ( 68.76, 61.20) -- ( 68.76, 55.20);

\path[draw=drawColor,line width= 0.4pt,line join=round,line cap=round] ( 95.20, 61.20) -- ( 95.20, 55.20);

\path[draw=drawColor,line width= 0.4pt,line join=round,line cap=round] (121.64, 61.20) -- (121.64, 55.20);

\path[draw=drawColor,line width= 0.4pt,line join=round,line cap=round] (148.08, 61.20) -- (148.08, 55.20);

\path[draw=drawColor,line width= 0.4pt,line join=round,line cap=round] (174.52, 61.20) -- (174.52, 55.20);

\path[draw=drawColor,line width= 0.4pt,line join=round,line cap=round] (200.95, 61.20) -- (200.95, 55.20);

\node[text=drawColor,anchor=base,inner sep=0pt, outer sep=0pt, scale=  1.00] at ( 68.76, 39.60) {200};

\node[text=drawColor,anchor=base,inner sep=0pt, outer sep=0pt, scale=  1.00] at ( 95.20, 39.60) {400};

\node[text=drawColor,anchor=base,inner sep=0pt, outer sep=0pt, scale=  1.00] at (121.64, 39.60) {600};

\node[text=drawColor,anchor=base,inner sep=0pt, outer sep=0pt, scale=  1.00] at (148.08, 39.60) {800};

\node[text=drawColor,anchor=base,inner sep=0pt, outer sep=0pt, scale=  1.00] at (174.52, 39.60) {1000};

\path[draw=drawColor,line width= 0.4pt,line join=round,line cap=round] ( 49.20, 65.14) -- ( 49.20,159.69);

\path[draw=drawColor,line width= 0.4pt,line join=round,line cap=round] ( 49.20, 65.14) -- ( 43.20, 65.14);

\path[draw=drawColor,line width= 0.4pt,line join=round,line cap=round] ( 49.20, 88.78) -- ( 43.20, 88.78);

\path[draw=drawColor,line width= 0.4pt,line join=round,line cap=round] ( 49.20,112.41) -- ( 43.20,112.41);

\path[draw=drawColor,line width= 0.4pt,line join=round,line cap=round] ( 49.20,136.05) -- ( 43.20,136.05);

\path[draw=drawColor,line width= 0.4pt,line join=round,line cap=round] ( 49.20,159.69) -- ( 43.20,159.69);

\node[text=drawColor,rotate= 90.00,anchor=base,inner sep=0pt, outer sep=0pt, scale=  1.00] at ( 34.80, 65.14) {0};

\node[text=drawColor,rotate= 90.00,anchor=base,inner sep=0pt, outer sep=0pt, scale=  1.00] at ( 34.80, 88.78) {50};

\node[text=drawColor,rotate= 90.00,anchor=base,inner sep=0pt, outer sep=0pt, scale=  1.00] at ( 34.80,112.41) {100};

\node[text=drawColor,rotate= 90.00,anchor=base,inner sep=0pt, outer sep=0pt, scale=  1.00] at ( 34.80,136.05) {150};

\node[text=drawColor,rotate= 90.00,anchor=base,inner sep=0pt, outer sep=0pt, scale=  1.00] at ( 34.80,159.69) {200};

\path[draw=drawColor,line width= 0.4pt,line join=round,line cap=round] ( 49.20, 61.20) --
	(220.52, 61.20) --
	(220.52,167.61) --
	( 49.20,167.61) --
	( 49.20, 61.20);
\end{scope}
\begin{scope}
\path[clip] (  0.00,  0.00) rectangle (245.72,216.81);
\definecolor{drawColor}{RGB}{0,0,0}

\node[text=drawColor,anchor=base,inner sep=0pt, outer sep=0pt, scale=  1.20] at (134.86,188.07) {\bfseries Iraq war};

\node[text=drawColor,anchor=base,inner sep=0pt, outer sep=0pt, scale=  1.00] at (134.86, 15.60) {Downloaded URLs};

\node[text=drawColor,rotate= 90.00,anchor=base,inner sep=0pt, outer sep=0pt, scale=  1.00] at ( 10.80,114.41) {Relevance (accumulated)};
\end{scope}
\begin{scope}
\path[clip] ( 49.20, 61.20) rectangle (220.52,167.61);
\definecolor{drawColor}{RGB}{0,0,0}

\path[draw=drawColor,line width= 1.2pt,dash pattern=on 4pt off 4pt ,line join=round,line cap=round] ( 55.55, 74.10) --
	( 68.76, 81.25) --
	( 81.98, 86.30) --
	( 95.20, 94.29) --
	(108.42,102.61) --
	(121.64,111.74) --
	(134.86,116.72) --
	(148.08,126.36) --
	(161.30,137.77);
\definecolor{drawColor}{RGB}{255,0,0}

\path[draw=drawColor,line width= 1.2pt,dash pattern=on 1pt off 3pt ,line join=round,line cap=round] ( 55.55, 68.53) --
	( 68.76, 69.48) --
	( 81.98, 69.82) --
	( 95.20, 71.56) --
	(108.42, 76.60) --
	(121.64, 82.98) --
	(134.86, 88.32) --
	(148.08, 91.69) --
	(161.30, 94.83) --
	(174.52, 95.64);
\definecolor{drawColor}{RGB}{0,205,0}

\path[draw=drawColor,line width= 1.2pt,dash pattern=on 1pt off 3pt on 4pt off 3pt ,line join=round,line cap=round] ( 55.55, 74.89) --
	( 68.76, 84.35) --
	( 81.98, 92.71) --
	( 95.20,101.80) --
	(108.42,109.97) --
	(121.64,113.28) --
	(134.86,123.30) --
	(148.08,132.44) --
	(161.30,141.42) --
	(174.52,148.59) --
	(187.73,154.43) --
	(200.95,163.67);
\definecolor{drawColor}{RGB}{0,0,0}

\path[draw=drawColor,line width= 0.4pt,line join=round,line cap=round] ( 42.46,163.67) rectangle (111.12,115.67);
\definecolor{drawColor}{RGB}{255,0,255}

\path[draw=drawColor,line width= 1.2pt,line join=round,line cap=round] ( 49.66,154.07) -- ( 64.06,154.07);
\definecolor{drawColor}{RGB}{0,0,0}

\path[draw=drawColor,line width= 1.2pt,dash pattern=on 4pt off 4pt ,line join=round,line cap=round] ( 49.66,144.47) -- ( 64.06,144.47);
\definecolor{drawColor}{RGB}{255,0,0}

\path[draw=drawColor,line width= 1.2pt,dash pattern=on 1pt off 3pt ,line join=round,line cap=round] ( 49.66,134.87) -- ( 64.06,134.87);
\definecolor{drawColor}{RGB}{0,205,0}

\path[draw=drawColor,line width= 1.2pt,dash pattern=on 1pt off 3pt on 4pt off 3pt ,line join=round,line cap=round] ( 49.66,125.27) -- ( 64.06,125.27);
\definecolor{drawColor}{RGB}{0,0,0}

\node[text=drawColor,anchor=base west,inner sep=0pt, outer sep=0pt, scale=  0.80] at ( 71.26,151.31) {Unfocused};

\node[text=drawColor,anchor=base west,inner sep=0pt, outer sep=0pt, scale=  0.80] at ( 71.26,141.71) {C-F};

\node[text=drawColor,anchor=base west,inner sep=0pt, outer sep=0pt, scale=  0.80] at ( 71.26,132.11) {T-F};

\node[text=drawColor,anchor=base west,inner sep=0pt, outer sep=0pt, scale=  0.80] at ( 71.26,122.51) {CT-F};
\end{scope}
\end{tikzpicture}}
  \scalebox{.68}{\input{figures/wulff}}

  \vspace{-\baselineskip}
  \caption{Accumulated relevance of different event collections.}
  \label{fig:crawler-hr}
  \vspace{-\baselineskip}
\end{figure}

\subsection{Extraction Evaluation}

Our focused crawling approach allows us to adjust the effort invested into the
extraction by changing the number of documents processed.
By increasing this number to the size of the archive we could clearly guarantee
that this method finds all the relevant documents, as long as they are reachable through links.
However, the proposed approach should be able to extract most of the relevant
documents early on, so that the extraction can be stopped 
when not sufficiently many relevant documents are discovered anymore or 
when the user is satisfied with the collection.
We therefore look at the accumulated relevance (\ie the sum of the relevance values of the extracted documents) of the collected results as a function of crawl runtime.
extraction process.
Additionally, we look at the number of documents that the crawler attempts to capture but are missing from the archive.

The relevance of the extracted documents is computed with the \TopicFocus relevance function.
This is possible because we estimate the relevance of a document during the crawl using the content of a linking document and evaluate using the content of the actual document.
A small annotation experiment (omitted for space) showed that this relevance measure correlates with the actual relevance.

For each of the \events events we started a crawl using each of the
configurations described above.
Each crawl ran  until it had retrieved 100,000 documents or until the crawler queue was empty.
Fig.~\ref{fig:crawler-hr} shows the accumulated relevance of document collections for selected events in relation to the number of documents crawled.
This function should ideally start with a strong incline, meaning that the crawler fetches many relevant documents early on, flattening into a plateau when no relevant documents are available anymore.
We see that for all topics the \TopicFocus and \TopicTempFocus functions outperform the \TempFocus function and the unfocused baseline both in terms of average relevance of documents retrieved at any given point and total relevance.
The \TopicFocus function often performs slightly better than the \TopicTempFocus function, although closer analysis shows that the differences between both functions often result from discovering some highly relevant hosts earlier.

The relevance focused strategies manage to uncover more potentially relevant URLs even if they are not contained in the locally available Web archive.
This is shown by the number of URLs that each focusing method considers (see Table~\ref{tab:crawl-size}), where we see an increase in discovered URLs for these methods.
Based on this result, the development of methods for cross-archive collection extraction is an interesting direction for future research.

\begin{table}[t]
  \centering
  \caption{URLs considered for each event crawl for different relevance
  strategies.}
  \label{tab:crawl-size}
  \begin{tabular}{lrrr}
    \toprule
    Topic       &       \TopicTempFocus & Unfocused   & Ratio\\
    \midrule
    Costa Concordia grounding       &	239,628&	142,851&	1.67\\
    German federal election 2009  &	283,311&	161,934&	1.74\\
    Iraq War       &	1,862&	2,192&	0.84\\
    Pope Election 2013    & 	2,057&	1,624&	1.26\\
    Stuttgart 21 protests  &	2,070&	1,513&	1.36\\
    Resignation of President Wulff \       &	213,039&	149,706&	1.42\\
    \bottomrule
  \end{tabular}
\end{table}

\begin{table}[tb]
  \caption{Effect of temporal scope and keyword parameters. Each row shows improvement ratio of the accumulated relevance for a topic with event-specific time parameters (left) or keywords (right). The last line contains the average improvement over all topics. All values are statistically significant at $p=0.01$.}
  \fontsize{7}{9}\selectfont
  \centering
  \begin{tabular}{lrlrl|lrlrl}
    \toprule
    \textbf{Event} & \multicolumn{2}{c}{\textbf{Time}} & \multicolumn{2}{c|}{\textbf{Keywords}} &     \textbf{Event} & \multicolumn{2}{c}{\textbf{Time}} & \multicolumn{2}{c}{\textbf{Keywords}} \\
    & \TempFocus & \TopicTempFocus & \TopicFocus & \TopicTempFocus &     & \TempFocus & \TopicTempFocus & \TopicFocus & \TopicTempFocus \\
    \midrule
    Book by Thilo Sarrazin	& 0.98	& 0.99	& 1.28	& 1.07 	&               Iraq war    & 0.92  & 1.19  & 1.05  & 1.13 \\
    Eruption of Eyjafjallaj\"okull	& 0.99	& 1.20	& 0.83	& 0.88 	&       Launch of LHC       & 1.09  & 0.72  & 1.21  & 0.99 \\
    European Stability Mechanism	& 1.16	& 4.07	& 1.02	& 1.04 	&       Costa Concordia grounding  & 1.14  & 1.49  & 0.92  & 0.98 \\
    European floods 2013	& 1.12	& 1.12	& 1.39	& 1.49 	&               Loveparade disaster & 0.84  & 1.25  & 0.81  & 0.97 \\
    Eurovision Song Contest 2010	& 1.00	& 1.73	& 1.06	& 0.68 	&       NSU process & 1.01  & 1.24  & 1.05  & 1.05 \\
    Football World Cup 2006	& 0.58	& 1.27	& 1.23	& 1.10 	&               Olympia 2004   & 0.94  & 1.03  & 1.20  & 1.34 \\
    Football World Cup 2010	& 1.59	& 1.09	& 1.11	& 1.10 	&               Olympia 2008   & 1.27  & 1.48  & 1.39  & 1.50 \\
    Fukushima nuclear disaster	& 1.17	& 1.73	& 1.03	& 1.02 	&               Olympia 2012   & 1.18  & 1.11  & 1.16  & 1.12 \\
    German federal election 2002	& 1.21	& 1.48	& 1.35	& 1.02 	&       Olympia 2010   & 1.02  & 1.37  & 1.24  & 1.65 \\
    German federal election 2005	& 1.33	& 1.41	& 1.14	& 0.89 	&       Pope Election 2005  & 1.17  & 1.08  & 1.10  & 1.09 \\
    German federal election 2009	& 1.27	& 1.84	& 1.03	& 0.96 	&       Pope Election 2013  & 1.07  & 1.50  & 0.99  & 0.95 \\
    German federal election 2013	& 1.12	& 2.17	& 0.84	& 0.92 	&       Resig. of Pres. Wulff      & 1.03  & 1.05  & 1.00  & 1.03 \\
    Guttenberg plagiarism affair	& 0.96	& 1.01	& 1.24	& 1.19 	&       Snowden leaks       & 1.46  & 1.43  & 1.18  & 1.19 \\
    \midrule
    \multicolumn{7}{l}{average} 1.10	& 1.44	& 1.10	& 1.08 \\
    \bottomrule
  \end{tabular}
  \vspace{-\baselineskip}
  \label{tab:parameters}
\end{table}

\subsection{Effect of the Temporal Scope Parameters}
\label{sec:eval-time-parameters}
In the \CS we require that the user specifies \emph{lead} and \emph{cool-down times} for the event (cf. Section~\ref{sec:temporal_relevance})
to adapt the temporal relevance function to different event types.
We crawled each event using a exponential decay function with a fixed decay and compared it to the crawl using the specified lead and cool-down times.
Table~\ref{tab:parameters} (left columns) shows the relevance improvement of the time-sensitive relevance functions over the corresponding baseline.
We see that the event-specific parameters cause an improvement for most of the events.
On average this improvement is moderate, but statistically significant.

\subsection{Effect of Keywords in the Specification}
\label{sec:effect-content-keywo}

We use the keywords in the \CS to clarify the topical intent (cf. Section~\ref{sec:focussing-relevance}).
To measure the impact, we crawled using the same reference documents with and without keywords to describe the topical scope.
Table~\ref{tab:parameters} (right columns) shows
the relevance improvement  of the \TempFocus and \TopicTempFocus relevance functions
compared to the corresponding baseline.
We see that the addition of keywords leads on average to a statistically
significant improvement.
Some events such as the floods in Europe during 2013 can be better focused using keywords, whereas for other events adding keywords leads
to a small loss in effectiveness.
Further research is needed to better understand the
influence of keywords.

\section{Conclusions and Outlook}
\label{sec:conclusions-outlook}

In this work we presented a novel method to
create interlinked event-centric collections from large-scale Web archives.
The key of this method is to adapt focused Web crawling to previously collected Web archives and to select documents by iteratively following links from relevant documents.
We proposed relevance estimation functions that take the
temporal and topical aspects of the documents into account and evaluated them as part of the focused extraction process.
Specifically, we demonstrated that the relevance
function \TopicTempFocus can improve on topical content selection methods 
by taking temporal information into account.
This holds especially for events that occur repeatedly in similar form, such as Olympic games or elections, where the different instances are hard to distinguish using only topical information.
We showed that our re-crawling method can retrieve event-centric collections from large-scale Web archives, especially using the \TopicTempFocus relevance function, and discussed how the method deals with the challenges inherent to Web archives.

Our method presents a first step towards the extraction of event-centric collections.
Further research is needed to understand the influence of extraction methods, relevance functions and parameters in regard to different events, time periods and Web archives.
For Web archives that have full-text indexes, methods based on full-text search should be investigated.
Furthermore, cross-archive collection extraction is an interesting direction for future research.
We therefore provide our source code and evaluation data to encourage similar efforts
\footnote{\url{https://github.com/gerhardgossen/archive-recrawling}}.

\section*{Acknowledgments}
\small{This work was partially funded by the ERC under ALEXANDRIA (ERC 339233), H2020 under SoBigData (RIA 654024) and BMBF under Data4UrbanMobility (02K15A040).}

\bibliographystyle{splncs03}
\bibliography{dl}

\end{document}